\documentclass[conference]{IEEEtran}

\ifCLASSINFOpdf

\else

\fi

\usepackage{amssymb}
\usepackage{graphicx}
\usepackage{subfigure}
\usepackage{bm}
\usepackage{booktabs}
\usepackage{CJK}
\usepackage{indentfirst}
\usepackage{amsmath}
\begin{document}

\title{Model-Driven Channel Estimation for OFDM Systems Based on Image Super-Resolution Network}

\author{\IEEEauthorblockN{Xin Ru, Li Wei, and Youyun Xu}
\IEEEauthorblockA{National Engineering Research Center of Communications $\&$ Networking\\Nanjing University of Posts and Telecommunications\\Nanjing 210003, China\\Email: \{1018010132, weili, yyxu\}@njupt.edu.cn}}

\maketitle

\begin{abstract}
In this paper, we propose a model-driven channel estimation method utilizing a convolutional neural network (CNN) derived from image super-resolution (SR). Instead of completely abandoning traditional communication modules as data-driven approaches always do, we first obtain rough channel matrix by the simplest least square (LS) channel estimation, and then use a CNN-based SR network to learn the subtle connections within channel matrix for accuracy improvement. The results show that our proposed method significantly outperforms the linear minimum mean squared error (LMMSE) estimation and has potential in spectrum saving. The impacts of network structure and SNR mismatch problem are also investigated in this paper.

\end{abstract}

\begin{IEEEkeywords}
Deep learning, channel estimation, image super-resolution, OFDM.
\end{IEEEkeywords}

\IEEEpeerreviewmaketitle

\section{Introduction}
Channel estimation, mainly designed for obtaining channel state information (CSI), is an integral part of OFDM systems. To a great extent, the accuracy of estimated CSI determines the performance of channel equalization in resisting the effects of multipath channel fading. Besides, power distribution and resource scheduling at the transmitter also need the assistance of the CSI feedback. Among traditional channel estimation methods, the minimum mean-squared error (MMSE) \cite{663750} estimation is known to be the optimal algorithm which suffers serious burden of computational complexity. And the least square (LS) \cite{6024619} estimation is simple but achieves poor performance. In practical applications, modified algorithms of the above two methods are more widely used, which usually make a trade-off between complexity and estimation accuracy according to actual needs, such as linear MMSE estimation \cite{504981} and low-rank approximation of LMMSE \cite{701321}. Those modifications of MMSE estimation inevitably introduce estimation errors since lots of channel information has been abandoned, and the results demonstrate the performance floor in high SNR. It is difficult to built a tractable mathematical models with superior performance when taking all the channel effects into account.
\par Recently, the remarkable ability of deep learning (DL) in characterizing sophisticated natural images or language inspires researchers to rethink the communication system architecture. Different from traditional mathematical modeling, DL-based methods are mainly based on the neural networks (NN) rather than humans to learn the processes of communication system, due to which some complicated features can be extracted effectively through powerful learning ability of NN. According to the different ways of integration, the existing DL-based OFDM channel estimation can be generally divided into two categories, data-driven approaches and model-driven approaches.
\par In data-driven approaches, channel estimation is completely realized by a trained network without any traditional estimation algorithms. For example, a DL-based method proposed in \cite{8052521} replaces the whole OFDM receiver with a fully connected network, which is the most straightforward way to reconstruct communication systems. In this method, the receiver is regarded as a black box, and channel estimation is jointly optimized with other modules, such as signal detection and demodulation. However, specific CSI cannot be obtained from this receiver because of the prohibitive difficulties in distinguishing the functions of each network layer. Moreover, data-driven approaches are limited by the need for huge amounts of training datasets and long training time to reach convergence \cite{8715338}.
\par In contrast to data-driven approaches, model-driven approaches preserve the basic OFDM system framework and focuse on the optimization or compensation for traditional algorithms. Aided by expert knowledge, it is much easier to obtain specific CSI and achieve faster convergence speed with less training datasets. The ComNet, proposed in \cite{Gao2018ComNet}, is the typical representative of model-driven approaches. In this method, the OFDM receiver is composed of two subnets in the manner of cascading modules similar to traditional communication system. The channel estimation subnet is initialized by LS estimation and applied to refine the coarse received signal. And the numerical results demonstrate that the ComNet outperforms the data-driven method in \cite{8052521}.
\par However, the fully connected network used in the ComNet involves large-scale tunable parameters leading to rapidly increasing computational burden when the network size grows. And the whole network has to be retrained once the size of input datasets changes \cite{8751143}. To solve those problems, convolutional neural network (CNN) is one of the effective solutions, which is benefited from  the concepts of local connectivity and weight sharing. But it is a challenging work to find the most suitable CNN for channel estimation, because the structure of CNN is much more flexible than that of fully connected network. In \cite{Soltani2018DeepLC}, two CNN-based image processing networks are cascaded to refine the channel time-frequency response.
\par Inspired by this idea, we propose a model-driven channel estimation method that incorporates image super-resolution (SR) network into the traditional estimator. In this method, the channel matrix is firstly acquired by the LS estimation, and then sent into the fast super-resolution convolutional neural network (FSRCNN) \cite{10.1007/978-3-319-46475-6_25} for estimation accuracy improvement.$\footnote{FSRCNN is a CNN-based SR network that learns the mapping between low-resolution images and high-resolution images.}$ Since the FSRCNN is totally based on CNN and performs directly on the low-resolution images, our proposed FSRCNN-based channel estimation (FSRCE) method can be applied to various systems with different input sizes. From the simulation results, the FSRCE is able to provide more accurate channel estimation than the LMMSE and the method in \cite{Soltani2018DeepLC}, even with fewer pilots inserted. We also investigate the MSE performance of the proposed method with some structural adjustments to achieve the trade-off between estimation accuracy and network complexity.

\section{TRADITIONAL ALGORITHMS}
\par Transmitted through a frequency selective multipath fading channel, the signal at the receiver can be represented as \begin{equation}
\emph{Y}(k)=\emph{H}(k)\emph{X}(k)+\emph{W}(k), k=0,1,\ldots,\emph{N}-1 \label{1}
\end{equation}where $\emph{X}(k)$ and $\emph{Y}(k)$ denote the $N$-point FFT of transmitted signal and received signal, respectively. $\emph{H}(k)$ refers to channel frequency response, and $\emph{W}(k)$ represents addictive white Gaussian noise with zero-mean and variance $\sigma^2$. In this paper, we consider the comb-type pilot arrangement. Each OFDM symbol consists of $\emph{N}_d$ subcarriers for data and $\emph{N}_p$ subcarriers for pilots, and there are $\emph{N}_d+\emph{N}_p=N$.
\par The simplest algorithm to estimate the channel is LS estimation \cite{6024619} \begin{equation}
\hat{\emph{H}}_\emph{LS}=arg\,min\{(\emph{Y}-\emph{X}\emph{H}_\emph{LS})^H(\emph{Y}-\emph{X}\emph{H}_\emph{LS})\} \label{2}
\end{equation}where $(\cdot)^H$ denotes the Hermitian of the matrix, $\emph{H}_\emph{LS}$ is the estimated channel. With the aid of pilots, the estimated result of (2) in pilot position is \begin{equation}
\hat{\emph{H}}_{\emph{LS},p}=\frac{\emph{Y}_p}{\emph{X}_p} \label{3}
\end{equation}where $\emph{Y}_p$ and $\emph{X}_p$ are the received and transmitted pilots, $\hat{\emph{H}}_{\emph{LS},p}$ is estimated CSI at pilot subcarriers. Then, channel state information at the data subcarriers $\hat{\emph{H}}_{\emph{LS},d}$ can be obtained by interpolating $\hat{\emph{H}}_{\emph{LS},p}$ in a linear or spline way.
\par The major drawback of LS estimation is its sensitivity to noise and inter-carrier interference, which seriously limits its performance. If there are high estimation performance requirement and sufficient computational redundancy, MMSE estimation is a better choice \cite{663750} \begin{equation}
\hat{\emph{H}}_{\emph{MMSE},p}=\emph{R}_{\emph{H}_p\emph{H}_p}(\emph{R}_{\emph{H}_p\emph{H}_p}+\sigma^2(\emph{X}_p\emph{X}_p^H)^{-1})^{-1}\hat{\emph{H}}_{\emph{LS},p} \label{4}
\end{equation}where $\emph{R}_{\emph{H}_p\emph{H}_p}=E\{\emph{H}_p\emph{H}_p^H\}$ is the autocorrelation matrix of channel in pilot positions. The matrix inversion $(\emph{X}_p\emph{X}_p^H)^{-1}$ has to be calculated every time $X_p$ changes. To solve this problem, a simplified linear MMSE estimation given in \cite{701321} replaces the matrix inversion $(\emph{X}_p\emph{X}_p^H)^{-1}$ by its expectation $E\{(\emph{X}_p\emph{X}_p^H)^{-1}\}=E\{|1/X_p(k)|^2\}I$. Thus, the linear MMSE estimation is obtaioned as \begin{equation}
\hat{\emph{H}}_{\emph{LMMSE},p}=\emph{R}_{\emph{H}_p\emph{H}_p}(\emph{R}_{\emph{H}_p\emph{H}_p}+\frac{\beta}{\emph{SNR}} \emph{I})^{-1}\hat{\emph{H}}_{\emph{LS},p} \label{5}
\end{equation}where $\beta=E\{|\emph{X}_p(k)|^{2}\}E\{|1/\emph{X}_p(k)|^2\}$ is a constant depending on the signal constellation, and $SNR=E\{|X_p(k)|^2\}/\sigma_n^2$, $\sigma_n^2$ is the variance of $W(k)$. The complexity can be further reduced if $\emph{R}_{\emph{H}_p\emph{H}_p}$ and $SNR$ are both known beforehand so that the term $\emph{R}_{\emph{H}_p\emph{H}_p}(\emph{R}_{\emph{H}_p\emph{H}_p}+\frac{\beta}{\emph{SNR}}\emph{I})^{-1}$ only needs to be calculated once.
\par Apparently, the autocorrelation of channel matrix plays an essential role in pilot-aided channel estimation. In other words, the features between the channel matrix elements in pilot positions are the key foundation for obtaining the whole channel response. Similarly, image super-resolution (SR) is a typical technique in computer vision to improve image resolution by filling new pixels, which is mainly based on learning the subtle changes between adjacent pixels. Therefore, the method of improving image resolution can be analogous to the process of obtaining the entire channel response utilizing the channel information in pilot positions.

\section{SR-BASED CHANNEL ESTIMATION}

\subsection{SR-based CE Architecture}
\begin{figure}
\centering
\setlength{\abovecaptionskip}{-0.1cm}   
\setlength{\belowcaptionskip}{-0.5cm}
\includegraphics[width=0.52\textwidth,height=0.18\textheight]{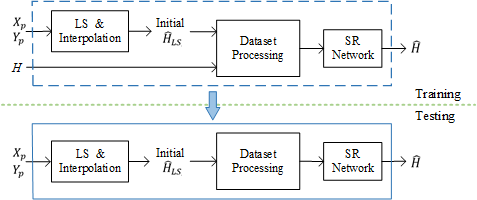}
\caption{SR-based channel estimation method}
\end{figure}

\begin{figure*}
\centering
\setlength{\abovecaptionskip}{-0.1cm}   
\setlength{\belowcaptionskip}{-0.5cm}
\includegraphics[width=0.9\textwidth,height=0.19\textheight]{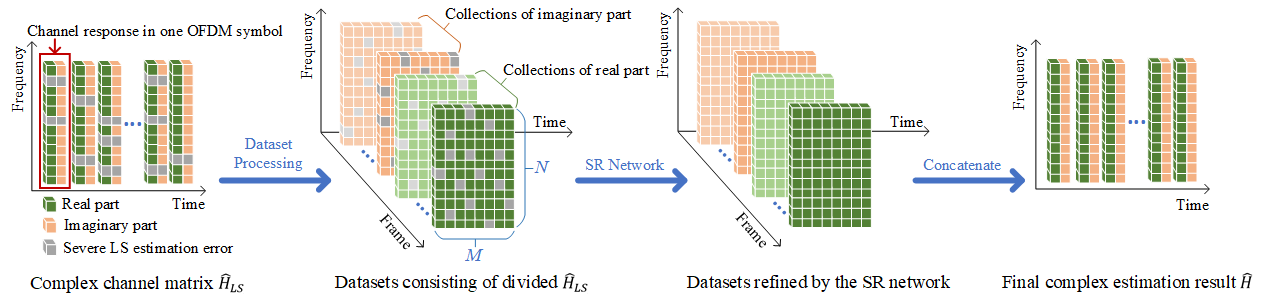}
\caption{Implementation of dataset processing and SR network}
\end{figure*}

As illustrated in Fig.1, there are two stages for the proposed method to acquire accurate channel estimation. In the training stage, the initial channel estimation at all subcarriers $\hat{\emph{H}}_{\emph{LS}}$ is roughly obtained by LS estimation and spline interpolation. After dataset processing, $\hat{\emph{H}}_{\emph{LS}}$ and $\emph{H}$ are sent into the SR network for training. The purpose of dataset processing is to obtain suitable dataset format for training. Detials about this module will be introduced in next subsection.
\par In the testing stage, we encapsulate the LS estimation, dataset proceessing module and the SR network into a whole SR-based channel estimator for online deployment. The output $\hat{\emph{H}}$ is the final estimation result. Testing datasets are generated in the same way as the training datasets to verify the mean-square error (MSE) of the model output. Additionally, the testing datasets follows the same statistical characteristics as the training datasets.
\par As for the structure of SR network, we adopt FSRCNN which achieves superior performance in SR than super-resolution convolutional neural network (SRCNN) \cite{Dong2016Image} used in \cite{Soltani2018DeepLC}. The structure of the two SR networks are listed as follows.
\begin{itemize}
 \item \textbf{SRCNN} consists of three parts. The first part is a patch extraction and representation layer with a $9\times9\times64$ filter which stands for 64 different $9\times9$ convolutional kernels. The second part is a non-linear mapping layer with a $1\times1\times32$ filter. And the last part is a reconstruction layer with a $5\times5\times1$ filter. The first two layers are both followed by a rectified linear unit (ReLU) as the activation function.
  \item \textbf{FSRCNN} consists of five parts. The first part is a feature extraction layer with a $5\times5\times56$ filter. The second part is a shrinking layer with a $1\times1\times12$ filter to reduce the feature dimension. And the third part is 4 non-linear mapping layers with the same filter size of $3\times3\times12$. The next part is an expanding layer with a $1\times1\times56$ filter. In \cite{10.1007/978-3-319-46475-6_25}, each of the above layers is activated by parametric rectified linear unit (PReLU) which is mainly used in training extremely deep networks. For the sake of computational efficiency, we still use ReLU as the activation function. The final part is a $9\times9\times1$ deconvolutional layer.
\end{itemize}

\subsection{Dataset Processing}
The dataset processing module is necessary for the proposed method. This is because our initial datasets are complex numbers, which is a commonly used representation in communications but not supported by current CNN. In this paper, we want to utilize the subtle relations between time slots of the multipath fading channel, so the format of datasets in both training and testing stages is a collection of $\emph{N} \times \emph{M}$ matrices, where $\emph{N}$ is the total number of subcarriers, and $\emph{M}$ is the number of OFDM symbols. In the simulation, one frame consists of M OFDM symbols. As illustrated in Fig.2, the complex channel matrix $\hat{H}_{LS}$ is a low-precision channel matrix obtained by LS estimation, in which the estimation error is ubiquitous but only the severe ones are marked in this figure. During the dataset processing, the real part and the imaginary part of $\hat{H}_{LS}$ are divided into multiple $\emph{N} \times \emph{M}$ matrices respectively. The collections of real and imaginary parts are analogous to low-resolution images, and then refined by the SR network. Finally, the high-resolution images are concatenated into complex estimation result $\hat{H}$. In our simulation, $N=64$ and $M=20$ are considered.

\subsection{Model Training}
In order to minimize the loss between the predicted channel matrix $\hat{\emph{H}}$ and the real channel matrix $\emph{H}$, we use MSE as the loss function \begin{equation}
\emph{L}(\theta)=\frac{1}{\emph{S}} \sum_{\emph{H} \in \mathbb{C}^{\emph{N}\times\emph{M}}} \| \hat{\emph{H}}-\emph{H} \|_2^2 \label{6}
\end{equation}where $\mathbb{C}$ is the collection of datasets, and $\emph{S}$ is the size of training datasets. The Adam \cite{article} optimizer is adopted to achieve better convergence. Furthermore, the batch size is 100 OFDM frames and the total epoch is set to be 800. The initial learning rate is 0.001 and decay 5-fold every 200 epochs. Meanwhile, validation datasets are also added during the training process to verify whether the loss function results increases as the training epoch becomes larger, i.e., to prevent the overfitting problem.

\section{SIMULATION RESULTS}
Simulations are conducted to investigate the performance of our proposed method. The entire training of the SR network is carried out under Tensorlow 1.14.0 and the datasets are collected from the simulation results of Matlab. For efficiency, we use Nvidia Quadro P4000 GPU to accelerate the training task. In the physical layer, we focus on SISO communication link with rayleigh fading multipath channel derived from widely used Jakes' model \cite{Zheng2011Improved}, where the carrier frequency is 2.6 GHz, the maximum delay is 16 sampling periods, and mobile velocity is 15 m/s. For the convenience of comparison later, we employ FSRCE as the concise convention of our proposed FSRCNN-based channel estimation method, and SRCE for the SRCNN-based method correspondingly.

\subsection{MSE Results}
\begin{figure}
\centering
\setlength{\abovecaptionskip}{-0.1cm}   
\setlength{\belowcaptionskip}{-1cm}
\includegraphics[width=0.45\textwidth,height=0.28\textheight]{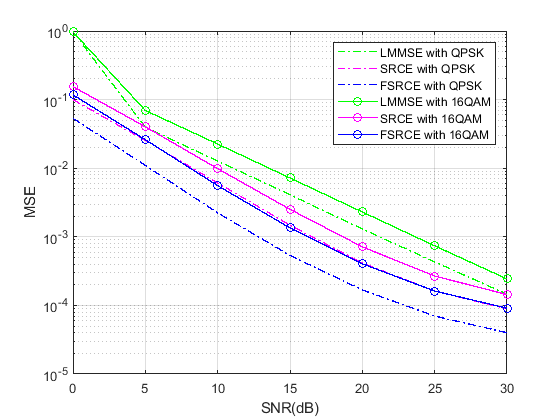}
\caption{MSE curves with QPSK and 16QAM modulation}
\end{figure}

The MSE performance of FSRCE, SRCE and the LMMSE baseline is depicted in Fig.3, where the number of pilots is 8. As shown in Fig.3, both deep learning based methods significantly outperform the benchmark over the whole SNR range, and the FSRCE shows the most superior performance. Interestingly, the MSE curve of the FSRCE under 16QAM modulation almost coincides with that of the SRCE under QPSK modulation, which means that FSRCE is able to achieve comparable performance with higher spectral efficiency. The results indicate that our proposed FSRCE can effectively extract valid features from wireless channel and compensate for the error of LS estimation results.

\subsection{Performance Analysis of Pilot Numbers}
\begin{figure}
\centering
\setlength{\abovecaptionskip}{-0.1cm}   
\setlength{\belowcaptionskip}{-1cm}
\subfigure[QPSK modulation]{
\includegraphics[width=0.45\textwidth,height=0.28\textheight]{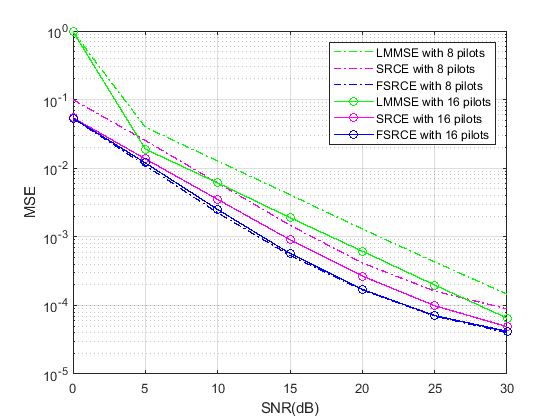}
}
\vspace{-0.1cm}  
\setlength{\abovecaptionskip}{-0.1cm}   
\setlength{\belowcaptionskip}{-1cm}
\subfigure[16QAM modulation]{
\includegraphics[width=0.45\textwidth,height=0.28\textheight]{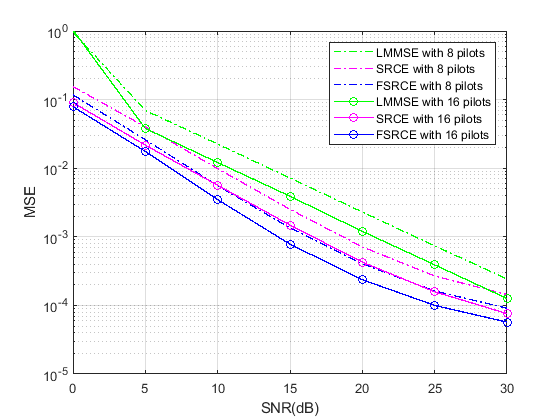}
}
\caption{MSE curves with different number of pilots}
\end{figure}

The impacts of different pilot numbers on the MSE performance are demonstrated in Fig.4. Compared to LMMSE and SRCE, FSRCE maintains competitive performance even with fewer pilots inserted. The results demonstrate the valuable potential of our proposed method in pilot saving.
\par It is worth noting that the performance gain between 8-pilot and 16-pilot FSRCE diminishes in QPSK case. The reason is that the signal with QPSK modulation is highly resistant to the effects of multipath channel and can be easily handled with 8 pilots, because of which the superiority of DL-based method becomes less obvious. And the compensation effects of our proposed method in QPSK modulation and 8 pilots case is close to the upper bound, so that simply adding pilots does not help much for further improvements.

\subsection{Network Structure Designing Analysis}
Generally, it is a sophisticated job to design a suitable structure of DL network for different situations. The 8-layer structure of FSRCNN that we mentioned in Section III-A is based on a series of testing work in \cite{10.1007/978-3-319-46475-6_25}, which does work well in image super-resolution but may not be the best choice for channel estimation. Given that the five parts of FSRCNN are indispensable, we make some changes to the third part, the non-linear mapping part.
\par In Fig.5, we compare the performance of FSRCE-x with LMMSE and SRCE, where 16QAM modulation and 16 pilots are adopted, and the suffix $-$x represents the number of non-linear mapping layers. It can be observed that the MSE performance of FSRCE-x does improve about 1 dB as the non-linear mapping layer increases from 2 to 4. However, when the layer increases to 6, the performance gain becomes negligible in high SNR and even diminishes in low SNR.
Although the learning ability of the network becomes stronger with the increase of layers, the difficulties of network training are also amplified exponentially. Standard backpropagation may fail to perform due to the local optima and other optimization challenges in the non-convex objective function as the network depth increases, which will lead to vanishing and exploding gradient problem \cite{8843652}. Therefore, the setting of the network layer number not only requires a trade-off between accuracy requirements and computational redundancy, but also needs to take practical implementation into consideration.

\begin{figure}
\centering
\setlength{\abovecaptionskip}{-0.1cm}   
\setlength{\belowcaptionskip}{-1cm}
\includegraphics[width=0.45\textwidth,height=0.28\textheight]{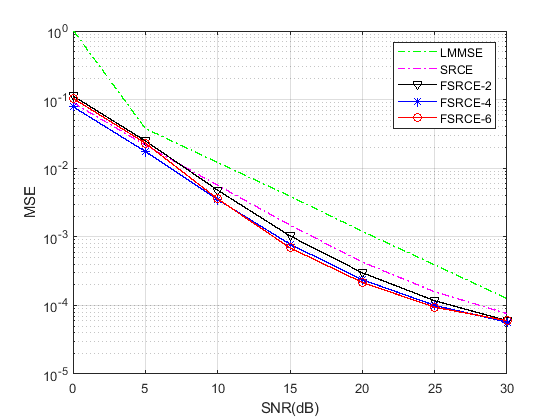}
\caption{MSE curves of FSRCE with different number of layers}
\end{figure}

\subsection{Impact of Training SNR Mismatch}
\begin{figure}
\centering
\setlength{\abovecaptionskip}{-0.1cm}   
\setlength{\belowcaptionskip}{-1cm}
\includegraphics[width=0.45\textwidth,height=0.28\textheight]{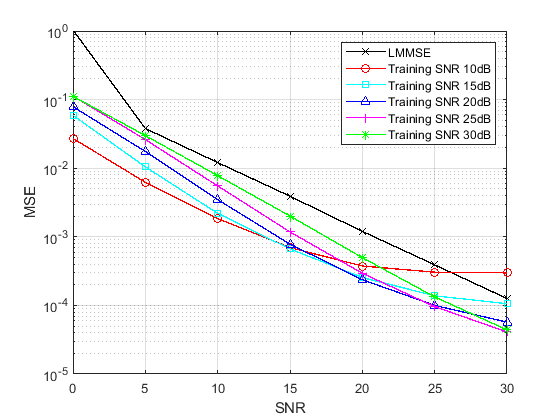}
\caption{MSE curves of FSRCE with different training SNR}
\end{figure}

In the above simulations, the datasets used for network training are generated in a fixed SNR (e.g. 20 dB in Fig.3-5), whereas the SNR is arbitrary in testing stage. Testing results of FSRCE under five different training SNRs are plotted in Fig.6 to investigate the impact of SNR mismatch, where 16 pilots and 16QAM modulation are adopted. It is obvious that the FSRCE can achieve the best MSE performance when training SNR is the same as testing SNR, and the performance degrades as the gap between the training and testing SNR becomes larger. The MSE performance loss caused by SNR mismatch is almost 5dB. In this case, we can take a compromise that choosing a fixed training SNR within a reasonable error range, or set a threshold to select the suitable network according to the actual channel conditions.
\section{CONCLUSION}
In this paper, a model-driven channel estimation method for OFDM systems is proposed. Comparisons with traditional LMMSE algorithm and existing CNN-based method indicate that the proposed method has competitive MSE performance and promising ability in spectrum saving. Furthermore, experiments in network structure designment and SNR mismatch are also conducted for the further performance evaluation of the proposed method.
\section*{Acknowledgment}
This paper is financially supported by the major research project of the National Natural Science Foundation of China (No. 61402520, No. 91438115, No. 61671252).

\bibliographystyle{IEEEtran}
\bibliography{bibfile}

\begin{thebibliography}{10}
\providecommand{\url}[1]{#1}
\csname url@samestyle\endcsname
\providecommand{\newblock}{\relax}
\providecommand{\bibinfo}[2]{#2}
\providecommand{\BIBentrySTDinterwordspacing}{\spaceskip=0pt\relax}
\providecommand{\BIBentryALTinterwordstretchfactor}{4}
\providecommand{\BIBentryALTinterwordspacing}{\spaceskip=\fontdimen2\font plus
\BIBentryALTinterwordstretchfactor\fontdimen3\font minus
  \fontdimen4\font\relax}
\providecommand{\BIBforeignlanguage}[2]{{%
\expandafter\ifx\csname l@#1\endcsname\relax
\typeout{** WARNING: IEEEtran.bst: No hyphenation pattern has been}%
\typeout{** loaded for the language `#1'. Using the pattern for}%
\typeout{** the default language instead.}%
\else
\language=\csname l@#1\endcsname
\fi
#2}}
\providecommand{\BIBdecl}{\relax}
\BIBdecl

\bibitem{663750}
{Meng-Han Hsieh} and {Che-Ho Wei}, ``Channel estimation for ofdm systems based
  on comb-type pilot arrangement in frequency selective fading channels,''
  \emph{IEEE Transactions on Consumer Electronics}, vol.~44, no.~1, pp.
  217--225, Feb 1998.

\bibitem{6024619}
A.~R. {James}, R.~S. {Benjamin}, S.~{John}, T.~M. {Joseph}, V.~{Mathai}, and
  S.~S. {Pillai}, ``Channel estimation for ofdm systems,'' in \emph{2011
  International Conference on Signal Processing, Communication, Computing and
  Networking Technologies}, July 2011, pp. 587--591.

\bibitem{504981}
J.~J. {van de Beek}, O.~{Edfors}, M.~{Sandell}, S.~K. {Wilson}, and P.~O.
  {Borjesson}, ``On channel estimation in ofdm systems,'' in \emph{1995 IEEE
  45th Vehicular Technology Conference. Countdown to the Wireless Twenty-First
  Century}, vol.~2, July 1995, pp. 815--819 vol.2.

\bibitem{701321}
O.~{Edfors}, M.~{Sandell}, J.~{van de Beek}, S.~K. {Wilson}, and P.~O.
  {Borjesson}, ``Ofdm channel estimation by singular value decomposition,''
  \emph{IEEE Transactions on Communications}, vol.~46, no.~7, pp. 931--939,
  July 1998.

\bibitem{8052521}
H.~{Ye}, G.~Y. {Li}, and B.~{Juang}, ``Power of deep learning for channel
  estimation and signal detection in ofdm systems,'' \emph{IEEE Wireless
  Communications Letters}, vol.~7, no.~1, pp. 114--117, Feb 2018.

\bibitem{8715338}
H.~{He}, S.~{Jin}, C.~{Wen}, F.~{Gao}, G.~{Ye Li}, and Z.~{Xu}, ``Model-driven
  deep learning for physical layer communications,'' \emph{IEEE Wireless
  Communications}, pp. 1--7, 2019.

\bibitem{Gao2018ComNet}
X.~Gao, S.~Jin, C.~K. Wen, and G.~Y. Li, ``Comnet: Combination of deep learning
  and expert knowledge in ofdm receivers,'' \emph{IEEE Communications Letters},
  vol.~22, no.~12, pp. 2627--2630, 2018.

\bibitem{8751143}
C.~{Fan}, X.~{Yuan}, and Y.~{Zhang}, ``Cnn-based signal detection for banded
  linear systems,'' \emph{IEEE Transactions on Wireless Communications},
  vol.~18, no.~9, pp. 4394--4407, Sep. 2019.

\bibitem{Soltani2018DeepLC}
M.~Soltani, V.~Pourahmadi, A.~Mirzaei, and H.~Sheikhzadeh, ``Deep
  learning-based channel estimation,'' \emph{IEEE Communications Letters},
  vol.~23, pp. 652--655, 2018.

\bibitem{10.1007/978-3-319-46475-6_25}
C.~Dong, C.~C. Loy, and X.~Tang, ``Accelerating the super-resolution
  convolutional neural network,'' in \emph{Computer Vision -- ECCV 2016},
  B.~Leibe, J.~Matas, N.~Sebe, and M.~Welling, Eds.\hskip 1em plus 0.5em minus
  0.4em\relax Cham: Springer International Publishing, 2016, pp. 391--407.

\bibitem{Dong2016Image}
C.~Dong, C.~C. Loy, K.~He, and X.~Tang, ``Image super-resolution using deep
  convolutional networks,'' \emph{IEEE Trans Pattern Anal Mach Intell},
  vol.~38, no.~2, pp. 295--307, 2016.

\bibitem{article}
D.~Kingma and J.~Ba, ``Adam: A method for stochastic optimization,''
  \emph{International Conference on Learning Representations}, 12 2014.

\bibitem{Zheng2011Improved}
Y.~R. Zheng and C.~Xiao, ``Improved models for the generation of multiple
  uncorrelated rayleigh fading waveforms,'' \emph{IEEE Communications Letters},
  vol.~6, no.~6, pp. 256--258, 2011.

\bibitem{8843652}
H.~H. {Tan} and K.~H. {Lim}, ``Vanishing gradient mitigation with deep learning
  neural network optimization,'' in \emph{2019 7th International Conference on
  Smart Computing Communications (ICSCC)}, June 2019, pp. 1--4.

\end{thebibliography}
\end{document}